\documentclass[runningheads]{svmult}
\usepackage{makeidx}
\usepackage{graphicx}
\usepackage{subeqnar}
\usepackage{multicol}

\begin{document}

\title*{Correlated Electrons in Carbon Nanotubes}
\toctitle{Correlated Electrons in Carbon Nanotubes}
\titlerunning{Correlated Electrons in Carbon Nanotubes}

\author{Arkadi A. Odintsov\inst{1} \and Hideo Yoshioka\inst{2}}
\authorrunning{A. Odintsov and H. Yoshioka}

\institute{Department of Applied Physics and DIMES,
Delft University of Technology, 2628 CJ Delft, The Netherlands \\
and Nuclear Physics Institute, Moscow State University,
Moscow 119899, Russia \\
\and Department of Physics, Nagoya University,
Nagoya 464-8602, Japan}

\maketitle

\begin{abstract}
Single-wall carbon nanotubes are almost ideal systems for the investigation
of exotic many-body effects due to non-Fermi liquid behavior of interacting
electrons in one dimension. Recent theoretical and experimental results are
reviewed with a focus on electron correlations. Starting from a microscopic
lattice model we derive an effective phase Hamiltonian for conducting
single-wall nanotubes with arbitrary chirality. The parameters of the
Hamiltonian show very weak dependence on the chiral angle, which makes the
low-energy physics of conducting nanotubes universal. The
temperature-dependent resistivity and frequency-dependent optical
conductivity of nanotubes with impurities are evaluated within the
Luttinger-like model. Localization effects are studied. In particular, we
found that intra-valley and inter-valley electron scattering can not coexist
at low energies. Low-energy properties of clean nanotubes are studied beyond
the Luttinger liquid approximation. The strongest Mott-like electron
instability occurs at half filling. In the Mott insulating phase electrons
at different atomic sublattices form characteristic bound states. The energy
gaps of $0.01-0.1$ eV occur in all modes of elementary excitations. 
We finally discuss observability of the Mott insulating phase in
transport experiments. The accent is made on the charge transfer from
external electrodes which results in a deviation of the electron density
from half-filling.
\end{abstract}

\section{Introduction}

Single-wall carbon nanotubes (SWNTs) are cylindrical fullerene structures
with diameters in the nanometer range and lengths of few micrometers \cite
{DekkerRev}. Experimental demonstration of electron transport through SWNTs 
\cite{Tans,Bockrath} has been followed by observations of atomic structure 
\cite{Wildoer,Odom,Johnson}, one-dimensional van Hove singularities \cite
{Wildoer,Odom}, standing electron waves \cite{Venema} and electron
correlations \cite{BockrathLL,SchonenbergerLL,Yao} in these systems.
Moreover, first prototypes of SWNT functional devices - diodes \cite
{Yao,Fuhrer} and field effect transistors \cite{Tans3} - have been
demonstrated recently.

On one hand, SWNTs can be viewed as giant macromolecules whose properties
can be learned from the first principle calculations. On the other hand,
SWNTs are perfect one-dimensional (1D) model systems to be studied by
methods of the solid state theory. This somewhat reductionist but insightful
approach provides a reasonable description for the bulk of experimental data
obtained up to date.

On a single-particle level physical properties of SWNTs are determined by
their geometry. Depending on the wrapping vector, SWNTs can either be 1D
metals or semiconductors with the energy gap in sub-electronvolt range. This
has been confirmed by direct observation of 1D van Hove singularities in
scanning tunneling microscopy experiments \cite{Wildoer,Odom}.

In this paper we address the role of the Coulomb interaction in 1D SWNTs and
review recent results on electron correlation effects. Away from
half-filling electron correlations are well described by Luttinger-like
models \cite{KBF,Egger-Gogolin,YOprl99}. In particular, the non-Fermi liquid
ground state of the system is characterized by a power-law suppression of
the density of electronic states near the Fermi level. This effect has been
observed in single- \cite{BockrathLL} and, presumably, multi-wall \cite
{SchonenbergerLL} nanotubes, as well as in junctions between metallic SWNTs 
\cite{Yao}. Transport properties of metallic nanotubes with impurities are
affected by the Coulomb interaction as well \cite{Yoshioka-impurities}.

The low-energy properties of SWNTs are different at half-filling due to the
umklapp scattering. The latter is coupled to the strongly interacting total
charge mode. This makes the umklapp scattering a strongly relevant
perturbation. As a result, the Mott-like electron instability occurs in the
electronic spectrum of SWNTs and energy gaps open in all modes of the
elementary excitations \cite{KBF,YOprl99}. The transition to the Mott
insulating phase should manifest itself by an increased resistivity of
half-filled SWNTs at low temperatures.

The plan of the paper is as follows. In Section 2 we derive an effective
low-energy model for metallic SWNTs with arbitrary chirality, which
naturally includes one-electron backscattering on impurities as well as
two-electron umklapp scattering at half-filling. We discuss the non-Fermi
liquid properties of clean SWNTs within the Luttinger-like model in Section
3 and evaluate electron transport in SWNTs with impurities in Section 4. The
effect of interactions beyond the Luttinger model is analyzed within the
renormalization group scheme in Section 5. The Mott-like electron
instability is further investigated in Section 6 using the self-consistent
harmonic approximation. Finally, we discuss observability of the
Mott-insulating phase in experiments (Section 7).

\section{Universal Model of Metallic Nanotubes}

\subsection{Microscopic theory}

\begin{figure}[tb]
\begin{center}
\includegraphics[width=.5\textwidth]{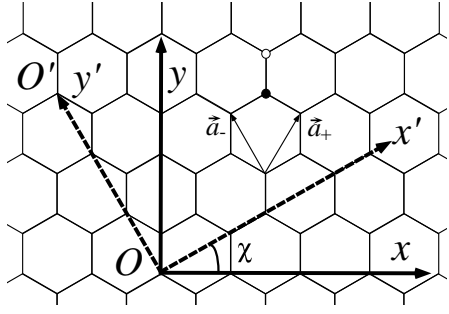}
\end{center}
\caption{Graphite lattice consists of two atomic sublattices $p=+,-$ denoted
by filled and open circles. SWNT at the angle $\protect\chi $ to $x$ axis
can be formed by wrapping the graphite sheet along $\vec{w}=OO^{\prime}$
vector. }
\label{figchir}
\end{figure}

Structurally uniform SWNTs can be characterized by the wrapping vector $\vec{%
w}=N_{+}\vec{a}_{+}+N_{-}\vec{a}_{-}$ given by the linear combination of
primitive lattice vectors $\vec{a}_{\pm }=(\pm 1,\sqrt{3})a/2$, with $%
a\approx 0.246$ nm (Fig. \ref{figchir}). It is natural to separate
non-chiral armchair ($N_{+}=N_{-}$) and zig-zag ($N_{+}=-N_{-}$) nanotubes
from their chiral counterparts. Recent scanning tunneling microscopy studies 
\cite{Wildoer,Odom} have revealed that individual SWNTs are generally
chiral. According to the single-particle model, the nanotubes with $%
N_{+}-N_{-}=0$ mod $3$ have gapless energy spectrum and are therefore
conducting; otherwise, the energy spectrum is gapped and SWNTs are
insulating.

We consider metallic SWNT whose axis $x^{\prime }$ forms an angle $\chi
=\arctan [(N_{-}-N_{+})/\sqrt{3}(N_{+}+N_{-})]$ with the direction of the
chains of carbon atoms ($x$ axis in Fig. \ref{figchir}). We expand the
standard single-particle Hamiltonian \cite{Wallace} $H_{k}$ for electrons on
two atomic sublattices $p=\pm $ of a graphite sheet (Fig. \ref{figchir})
near the Fermi points $\alpha \vec{K}$ (with the valley index $\alpha =\pm $%
, and $\vec{K}=(4\pi /3a,0)$) to the lowest order in $\vec{q}=\vec{k}-\alpha 
\vec{K}=q(\cos \chi ,\sin \chi )$. Introducing slowly varying Fermi fields $%
\psi _{p\alpha s}(x^{\prime })=L^{-1/2}\sum_{q=2\pi n/L}e^{iqx^{\prime
}}a_{p\alpha s}(\vec{q}+\alpha \vec{K})$, we obtain, 
\begin{equation}
H_{k}=-iv\sum_{p\alpha s}\alpha e^{-ip\alpha \chi }\int \mathrm{d}x^{\prime
}\psi _{p\alpha s}^{\dagger }\partial _{x^{\prime }}\psi _{-p\alpha s},
\label{Hk}
\end{equation}
where $v\approx 8.1\times 10^{5}$ m/s is the Fermi velocity, and $s=\pm $ is
the electron spin.

The kinetic term can be diagonalized by the unitary transformation 
\begin{equation}
\psi _{p\alpha s}=\frac{1}{\sqrt{2}}e^{-ip\alpha \chi /2}\sum_{r=\pm
}(r\alpha )^{\frac{1-p}{2}}\varphi _{r\alpha s}  \label{Utrans}
\end{equation}
to the basis $\varphi _{r\alpha s}$ of right- ($r=+$) and left- ($r=-$) and
moving electrons.

The Coulomb interaction has the form 
\begin{equation}
H_{int}=\frac{1}{2}\sum_{pp^{\prime },\left\{ \alpha _{i}\right\}
,ss^{\prime }}V_{pp^{\prime }}(2\bar{\alpha}K)\int dx^{\prime }\psi
_{p\alpha _{1}s}^{\dagger }\psi _{p^{\prime }\alpha _{2}s^{\prime
}}^{\dagger }\psi _{p^{\prime }\alpha _{3}s^{\prime }}\psi _{p\alpha _{4}s},
\label{Hint}
\end{equation}
with the matrix elements $V_{pp^{\prime }}(2\bar{\alpha}K)$ corresponding to
the amplitudes of intra- ($p=p^{\prime }$) and inter- ($p=-p^{\prime }$)
sublattice intra- ($\bar{\alpha}=0$) and inter- ($\bar{\alpha}=\pm 1$)
valley scattering (here $\bar{\alpha}=(\alpha _{1}-\alpha _{4})/2=(\alpha
_{3}-\alpha _{2})/2$). In Eq. (\ref{Hint}) we assume that the fields $\psi $
are varying slowly on the scale of the screening radius $R_{s}$ of the
Coulomb interaction determined by the distance to a gate electrode. This
corresponds to a contact-interaction approximation. Equation (\ref{Hint}) is
therefore valid at large length scale $x^{\prime }\gg R_{s}$ and low
electron energy $E\ll \hbar v/R_{s}$.

The dominant contribution to the intra-valley scattering amplitudes $%
V_{pp^{\prime }}(0)$ comes from the long range component of the Coulomb
interaction, $V_{pp}(0)\simeq V_{p-p}(0)\simeq e^{2}/C\simeq (2e^{2}/\kappa
)\ln (R_{s}/R)$, $C$ being the capacitance of SWNT per unit length \cite{KBF}%
. The differential part $\Delta V(0)=V_{pp}(0)-V_{p-p}(0)$ of intra-valley
scattering as well as the intra-sublattice inter-valley $V_{pp}(2K)$ are
estimated at $\Delta V(0),V_{pp}(2K)$ $\sim ae^{2}/\kappa R$ . Despite $%
\Delta V(0),V_{pp}(2K)$ being much smaller than $V_{pp}(0),$ they cause
non-Luttinger terms in the low-energy Hamiltonian which will be important in
the further analysis. The matrix elements $V_{pp^{\prime }}(2\bar{\alpha}K)$%
\ have been evaluated numerically for all SWNTs with radii $R$ in the range $%
2R/a=4-7$ ($2R/a=5.5$ for (10,10) SWNTs). We found that dimensionless
amplitudes $2\pi \kappa R[\Delta V(0),V_{pp}(2K)]/ae^{2}$ show very weak
dependence on the radius of SWNT and its chiral angle (see Table 1). The
results are sensitive to the value the short distance cutoff $a_{0}\sim a$
of the Coulomb interaction. 
\begin{table}[tb]
\caption{Scattering amplitudes $\Delta V(0)$, $V_{pp}(2K)$, $V_{p-p}(2K)$ in
units $ae^{2}/2\protect\pi \protect\kappa R$ for all SWNTs with $2R/a=4-7$.
The cutoff parameter $a_{0}=0.526a$ is obtained from the requirement that
the on-site interaction in the original tight-binding model is equal to the
difference between the ionization potential and electron affinity of $sp^{2}$
hybridized carbon \protect\cite{Moore}. }
\label{tab:ampl}\vspace{0.4cm}
\par
\begin{center}
\begin{tabular}{|c|c|c|l|}
\hline
$a_{0}/a$ & $\Delta V(0)$ & $V_{pp}(2K)$ & $|V_{p-p}(2K)|$ \\ \hline
$0.4$ & $0.44265-0.44274$ & $0.97060-0.97095$ & $0.6-2.2\times 10^{-3}$ \\ 
$0.526$ & $0.17378-0.17395$ & $0.53549-0.53561$ & $0.5-1.6\times 10^{-3}$ \\ 
$0.7$ & $0.04880-0.04895$ & $0.24778-0.24797$ & $0.3-1.5\times 10^{-3}$ \\ 
\hline
\end{tabular}
\end{center}
\end{table}

The inter-sublattice inter-valley scattering amplitude $V_{p-p}(2K)$ is
almost three orders of magnitude smaller than $\Delta V(0)$, $V_{pp}(2K)$.
This is due to the $C_{3}$ symmetry of a graphite lattice ($V_{p-p}(2K)=0$
for a plane graphite sheet). The matrix elements $V_{p-p}(2K)$ are generally
complex due to the asymmetry of effective 1D inter-sublattice interaction
potential (the matrix elements are real for symmetric zig-zag and armchair
SWNTs). Let us note that after the unitary transformation (\ref{Utrans}) of
the Hamiltonian $H=H_{k}+H_{int}$, the chiral angle $\chi $ enters \textit{%
only} into the inter-sublattice inter-valley scattering matrix elements $%
V_{p-p}(2K)$. Due to the smallness of these matrix elements, the low-energy
properties of chiral SWNTs are expected to be virtually independent of the
chiral angle.

\subsection{Bosonization}

The transformed Hamiltonian $H$ can be bosonized by introducing the phase
representation of the Fermi fields \cite{Egger-Gogolin,YOprl99}, 
\begin{equation}
\varphi _{r\alpha s}=\frac{\eta _{r\alpha s}}{\sqrt{2\pi \tilde{a}}}\exp %
\left[ \mathrm{i}rq_{F}x^{\prime }+\frac{\mathrm{i}r}{2}\left\{ \theta
_{\alpha s}+r\phi _{\alpha s}\right\} \right] .  \label{psi_bosonized}
\end{equation}
The phase variables $\theta _{\alpha s},\phi _{\alpha s}$ \ are further
decomposed into symmetric $\delta =+$ and antisymmetric $\delta =-$ modes of
the charge $\rho $ and spin $\sigma $ excitations, $O_{\alpha s}=O_{\rho
+}+sO_{\sigma +}+\alpha O_{\rho -}+\alpha sO_{\sigma -}$, $O=\theta ,\phi $.
The bosonic fields satisfy the commutation relation, $[\theta _{j\delta
}(x_{1}),\phi _{j^{\prime }\delta ^{\prime }}(x_{2})]=\mathrm{i}(\pi /2)%
\mathrm{sign}(x_{1}-x_{2})\delta _{jj^{\prime }}\delta _{\delta \delta
^{\prime }}$. The Majorana fermions $\eta _{r\alpha s}$ are introduced to
ensure correct anticommutation rules for different species $r,\alpha ,s$ of
electrons, and satisfy $[\eta _{r\alpha s},\eta _{r^{\prime }\alpha ^{\prime
}s^{\prime }}]_{+}=2\delta _{rr^{\prime }}\delta _{\alpha \alpha ^{\prime
}}\delta _{ss^{\prime }}$. The quantity $q_{F}=\pi n/4$ is related to the
deviation $n$ of the average electron density from half-filling, and $\tilde{%
a}\sim a$ is the parameter of the exponential ultraviolet cutoff.

Neglecting the inter-sublattice inter-valley scattering we arrive at the
universal phase Hamiltonian of metallic SWNTs, 
\begin{eqnarray}
H &=&\sum_{j=\rho ,\sigma }\sum_{\delta =\pm }\frac{v_{j\delta }}{2\pi }\int 
\mathrm{d}x^{\prime }\left\{ K_{j\delta }^{-1}(\partial _{x^{\prime }}\theta
_{j\delta })^{2}+K_{j\delta }(\partial _{x^{\prime }}\phi _{j\delta
})^{2}\right\} +\frac{1}{2(\pi \tilde{a})^{2}}\int \mathrm{d}x^{\prime } 
\nonumber \\
&&\{[\Delta V(0)-V_{pp}(2K)][\cos (4q_{F}x^{\prime }+2\theta _{\rho +})\cos
2\theta _{\sigma +}-\cos 2\theta _{\rho -}\cos 2\theta _{\sigma -}] 
\nonumber \\
&&-\Delta V(0)\cos (4q_{F}x^{\prime }+2\theta _{\rho +})\cos 2\theta _{\rho
-}+\Delta V(0)\cos (4q_{F}x^{\prime }+2\theta _{\rho +})\cos 2\theta
_{\sigma -}  \nonumber \\
&&-{\Delta V(0)}\cos 2\theta _{\sigma +}\cos 2\theta _{\rho -}+{\Delta V(0)}%
\cos 2\theta _{\sigma +}\cos 2\theta _{\sigma -}  \nonumber \\
&&-{V}_{pp}{(2K)}\cos (4q_{F}x^{\prime }+2\theta _{\rho +})\cos 2\phi
_{\sigma -}+{V}_{pp}{(2K)}\cos 2\theta _{\sigma +}\cos 2\phi _{\sigma -} 
\nonumber \\
&&+{V}_{pp}{(2K)}\cos 2\theta _{\rho -}\cos 2\phi _{\sigma -}+{V}_{pp}{(2K)}%
\cos 2\theta _{\sigma -}\cos 2\phi _{\sigma -}\},  \label{Hbos}
\end{eqnarray}
$v_{j\delta }=v\sqrt{A_{j\delta }B_{j\delta }}$ and $K_{j\delta }=\sqrt{%
B_{j\delta }/A_{j\delta }}$ being the velocities and interaction parameters
for different modes $j,\delta $ of excitations. The parameters $A_{j\delta }$%
, $B_{j\delta }$ are given by $A_{\rho +}=1+[8\bar{V}(0)-\Delta
V(0)/2-V_{pp}(2K)]/2\pi v$, $A_{\nu \delta }=1-[\Delta V(0)/2+\delta
V_{pp}(2K)]/2\pi v$, $B_{\nu \delta }=1+\Delta V(0)/4\pi v$, with $\bar{V}%
(0)=[V_{pp}(0)+V_{p-p}(0)]/2$. The renormalization of the parameters $%
K_{j\delta }$, $v_{j\delta }$ by the Coulomb interaction is strongest in $%
\rho +$ mode. Assuming $\kappa =1.4$ \cite{Egger-Gogolin} $R=0.7$ nm, and $%
R_{s}=100$ nm we obtain $K_{\rho +}\simeq 0.2$. The interaction in the other
modes is weak: $K_{j\delta }=1+O(a/R)$.

Let us note that the Hamiltonian (\ref{Hbos}) has the same form as the phase
Hamiltonian for a two-leg Hubbard-type ladder \cite{Lin-Balents-Fisher},
provided that the difference in definitions of the fields $\theta _{j-}$ and 
$\phi _{j-}$ ($j=\rho ,\sigma $) in terms of densities of right- and
left-movers in two energy bands is taken into account.

\subsection{Impurity scattering}

Disorder in the atomic potential is described by the Hamiltonian \cite
{Ando-Nakanishi,Yoshioka-impurities}, $H_{imp}=\sum_{p\alpha \alpha ^{\prime
}s}\int dxV_{p\bar{\alpha}}(x)\psi _{p\alpha s}^{\dagger }\psi _{p\alpha
^{\prime }s}$. Here the impurity potential $V_{p}(x)$ at the sublattice $%
p=\pm $ is decomposed into intra-valley ($\bar{\alpha}\equiv (\alpha
^{\prime }-\alpha )/2=0$) and inter-valley ($\bar{\alpha}=\pm 1$) scattering
components. First, we transform the Hamiltonian $H_{imp}$ to the basis of
right- and left-movers ($r=\pm $), see Eq. (\ref{Utrans}). When the range of
the impurity potential is much larger than the lattice constant, the
backward scattering ($r\rightarrow -r$) is ineffective \cite{Ando-Nakanishi}%
. We consider the case of short range impurity potential and retain
backscattering terms in the Hamiltonian. The forward scattering is discarded
because it does not contribute to the transport.

In the limit of weak impurity potential, the interaction between the
electrons and the impurities can be parameterized by uncorrelated Gaussian
random fields, $\eta (x)$ and $\xi (x)$ expressing the intra-valley and the
inter-valley backward scattering, respectively. The fields satisfy $\langle
\eta (x)\eta (x^{\prime })\rangle _{imp}=D_{1}\delta (x-x^{\prime })$ and $%
\langle \xi (x)\xi ^{\ast }(x^{\prime })\rangle _{imp}=D_{2}\delta
(x-x^{\prime })$, where $\langle \cdots \rangle _{imp}$ is the
configurational average. The factors $D_{1}$ and $D_{2}$ are given by $%
v/\tau _{1}$ and $v/\tau _{2}$ with the scattering time $\tau _{1}$ ($\tau
_{2}$) due to the intra-valley (inter-valley) backscattering. The
Hamiltonian of impurities is given by 
\begin{equation}
H_{imp}=\int dx\eta (x)\sum_{r\alpha s}\psi _{r\alpha s}^{\dagger }\psi
_{-r\alpha s}+\int dx\left\{ \xi (x)\sum_{rs}\psi _{r+s}^{\dagger }\psi
_{-r-s}+h.c.\right\} .  \label{eqn:Himp-3}
\end{equation}
Note that the intra-valley (inter-valley) backward scattering is
parameterized by a real $\eta (x)$ (complex $\xi (x)$) field. The
Hamiltonian $H_{imp}=H_{imp}^{1}+H_{imp}^{2}$ is expressed in terms of the
phase variables as follows, 
\begin{eqnarray}
H_{imp}^{1} &=&\frac{\mathrm{i}\sigma _{z}}{2\pi \tilde{a}}\int dx\eta
(x)\sum_{r\alpha s}r\alpha \exp \left( -2\mathrm{i}rq_{F}x\right)  \nonumber
\label{eqn:Himp1-b} \\
&\times &\exp \left\{ -\mathrm{i}r(\theta _{\rho +}+s\theta _{\sigma
+}+\alpha \theta _{\rho -}+\alpha s\theta _{\sigma -})\right\} , \\
H_{imp}^{2} &=&\frac{-\mathrm{i}\sigma _{y}}{2\pi \tilde{a}}\int
dx\sum_{rs}r\exp \left\{ -\mathrm{i}r(2q_{F}x+\theta _{\rho +}+s\theta
_{\sigma +})\right\}  \nonumber \\
&\times &\left[ \xi (x)\exp \left\{ -\mathrm{i}(\phi _{\rho -}+s\phi
_{\sigma -})\right\} +h.c.\right] .  \label{eqn:Himp2-b}
\end{eqnarray}
with the Pauli matrices $\sigma _{y},$ $\sigma _{z}$ originating from the
Majorana representation of the Fermi fields (\ref{psi_bosonized}).

\section{Luttinger model limit}

The phase Hamiltonian (\ref{Hbos}) of clean metallic SWNT consists of a part
quadratic in bosonic fields describing scattering of electrons within the
same branch of the spectrum and the non-quadratic part describing
inter-branch scattering. In the Luttinger model limit one neglects the
differential part $\Delta V(0)$ of intra-valley scattering and the
inter-valley scattering $V_{pp^{\prime }}(2K)$. In this case the Hamiltonian
(\ref{Hbos}) reads,

\begin{equation}
H_{L}=\sum_{j=\rho ,\sigma }\sum_{\delta =\pm }\frac{v_{j\delta }}{2\pi }%
\int \mathrm{d}x^{\prime }\left\{ K_{j\delta }^{-1}(\partial _{x^{\prime
}}\theta _{j\delta })^{2}+K_{j\delta }(\partial _{x^{\prime }}\phi _{j\delta
})^{2}\right\} ,  \label{HLutt}
\end{equation}
with parameters $v_{j\delta }=v\sqrt{A_{j\delta }}$ and $K_{j\delta }=1/%
\sqrt{A_{j\delta }}$ determined by $A_{\rho +}=1+4\bar{V}(0)/\pi v$, $%
A_{j\delta }=1$ for $(j\delta )\neq (\rho +)$. The parts of the Hamiltonian (%
\ref{HLutt}) in the four sectors of excitations $(j\delta )$ are decoupled.
In particular, using the relations $\rho =(2/\pi )\partial _{x}\theta _{\rho
+}$, $j=(2v/\pi )\partial _{x}\phi _{\rho +}$ for the charge density $e\rho $
and the electric current $ej$ ($e>0$) in SWNT, the Hamiltonian (\ref{HLutt})
in the total charge $\rho +$ sector can be rewritten as follows, 
\begin{equation}
H_{\rho +}=\frac{1}{\nu }\int \mathrm{d}x^{\prime }\left[ \frac{1}{2}\rho
^{2}+\frac{1}{2}\left( \frac{j}{v}\right) ^{2}\right] +\frac{1}{2}\int 
\mathrm{d}x^{\prime }\mathrm{d}x^{\prime \prime }\rho (x^{\prime
})V(x^{\prime }-x^{\prime \prime })\rho (x^{\prime \prime }),
\label{Hrhoplus}
\end{equation}
(in Eq. (\ref{HLutt}) the contact interaction approximation $V(x)=\bar{V}%
(0)\delta (x)$ has been made). In Eq. (\ref{Hrhoplus}) $\nu =4/\pi v$ is the
density of electronic states in SWNT; the first term describes the kinetic
energy of right and left moving electrons and the second corresponds to the
Hartree part of the interaction.

Electronic properties of SWNTs in the Luttinger limit have been investigated
in e.g. Refs. \cite{KBF,Egger-Gogolin}. The density of electronic states
near the Fermi level is suppressed in a power-law fashion. This is a
signature of the orthogonality catastrophe which occurs due to the
non-fermionic (bosonic) nature of low-energy excitations in SWNTs. As a
result, the differential conductance in the tunneling regime shows a
power-law dependence on temperature, $dI/dV\propto T^{\alpha }$, for $eV\ll
T $ and voltage, $dI/dV\propto V^{\alpha }$, for $eV\gg T$. The exponent $%
\alpha $ is given by $\alpha =(K_{\rho +}^{-1}+K_{\rho +}-2)/8$ for the
tunneling from a metallic electrode into the ''bulk'' of SWNT, $\alpha
=(K_{\rho +}^{-1}-1)/4$ for the tunneling into the end of SWNT, and $\alpha
=(K_{\rho +}^{-1}-1)/2$ for the end-to-end tunneling in nanotube
heterojunctions \cite{Yao}. Moreover, scaled differential conductance, $%
(dI/dV)/T^{\alpha }$, is a universal function of the ratio $eV/T$ . The
collapse of the data taken at different temperatures to a single universal
curve provides a comprehensive check of validity of the Luttinger model for
SWNTs \cite{BockrathLL,Yao}.

\section{Effect of impurities}

In this section, the transport properties of metallic SWNTs with impurities
are studied in the Luttinger model limit \cite{Yoshioka-impurities}. The
dynamical conductivity $\sigma (\omega )$ is expressed via the memory
function $M(\omega )$ as follows \cite{memory}, 
\begin{equation}
\sigma (\omega )=\frac{-\mathrm{i}\chi (0)}{\omega +M(\omega )},\ M(\omega )=%
\frac{\left( \langle F;F\rangle _{\omega }-\langle F;F\rangle _{\omega
=0}\right) /\omega }{-\chi (0)},  \label{eqn:M-omega-1}
\end{equation}
where $\langle A;A\rangle _{\omega }\equiv -\mathrm{i}\int
dx\int_{0}^{\infty }dt\mathrm{e}^{(\mathrm{i}\omega -\eta )t}\langle \lbrack
A(x,t),A(0,0)]\rangle $ with $\eta \rightarrow +0$, $\langle \cdots \rangle $
denotes the thermal average with respect to the Hamiltonian $H=H_{L}+H_{imp}$%
, $F=[j,H]$ with $j$ being the current operator, and $\chi (0)=\langle
j;j\rangle _{\omega =0}=-4v/\pi $.

The temperature dependence of the resistivity is given by, 
\begin{equation}
\rho =\rho _{B0}\frac{\Gamma ^{2}((K_{\rho +}+3)/4)}{\Gamma ((K_{\rho
+}+3)/2)}\left( \frac{2\pi T}{\omega _{F}}\right) ^{(K_{\rho +}-1)/2},
\label{rhoimpurity}
\end{equation}
where $\rho _{B0}=(\pi /2)\sum_{i=1,2}(v\tau _{i})^{-1}$, is the resistivity
of the non-interacting system in the Born approximation, $\omega _{F}=v/{%
\tilde{a}}$, and $\Gamma (z)$ is the gamma function. It is remarkable that $%
\rho /\rho _{B0}$ is independent of the scattering strength. The repulsive
Coulomb interaction ($K_{\rho +}<1$) leads to an enhancement of the
resistivity at low temperatures. For typical nanotubes with $K_{\rho
+}\simeq 0.2$, the resistivity scales as $\rho \propto T^{-0.4}$.

The optical conductivity $\mathrm{Re}\sigma (\omega )$ behaves like $\omega
^{(K_{\rho +}-5)/2}$ at high frequencies $\omega \gg T$ and as $\omega
^{(1-K_{\rho +})/2}$ at low frequencies. The position of the peak in $%
\mathrm{Re}\sigma (\omega )$ is given by, 
\begin{equation}
\frac{\omega }{\omega _{F}}\sim \left[ \frac{2}{\omega _{F}}(\frac{1}{\tau
_{1}}+\frac{1}{\tau _{2}})\frac{\tan \left\{ \pi (1-K_{\rho +})/4\right\} }{%
\Gamma ((K_{\rho +}+3)/2)}\right] ^{2/(3-K_{\rho +})},
\end{equation}

We will further study localization of electrons which corresponds to the
pinning of the phase fields $\theta _{j\delta },\phi _{j\delta }$. The
localization is described within the renormalization group (RG) formalism.
RG equations can be derived by assuming scale invariance of correlation
functions along the lines of Ref. \cite{Giamarchi-Schulz}, 
\begin{eqnarray}
(\mathcal{D}_{1})^{\prime } &=&\left\{ 3-(K_{\rho +}+K_{\sigma +}+K_{\rho
-}+K_{\sigma -})/2\right\} \mathcal{D}_{1}, \\
(\mathcal{D}_{2})^{\prime } &=&\left\{ 3-(K_{\rho +}+K_{\sigma +}+K_{\rho
-}^{-1}+K_{\sigma -}^{-1})/2\right\} \mathcal{D}_{2}, \\
(K_{j+})^{\prime } &=&-\left( {\mathcal{D}_{1}}/{X_{1}}+{\mathcal{D}_{2}}/{%
X_{2}}\right) K_{j+}^{2}u_{j+}, \\
(u_{j+})^{\prime } &=&-\left( {\mathcal{D}_{1}}/{X_{1}}+{\mathcal{D}_{2}}/{%
X_{2}}\right) {K_{j+}u_{j+}^{2}}, \\
(K_{j-})^{\prime } &=&-\left( {\mathcal{D}_{1}}K_{j-}^{2}/{X_{1}}-{\mathcal{D%
}_{2}}/{X_{2}}\right) u_{j-}, \\
(u_{j-})^{\prime } &=&-\left( {\mathcal{D}_{1}}K_{j-}/{X_{1}}+{\mathcal{D}%
_{2}}K_{j-}^{-1}/{X_{2}}\right) {u_{j-}^{2}},
\end{eqnarray}
where $(\ )^{\prime }$ denotes $\mathrm{d}/\mathrm{d}\ell $ with $\mathrm{d}%
\ell =\mathrm{d}\ln (\tilde{a}/a)$ ($\,\tilde{a}$ is the new lattice
constant), $X_{1}=u_{\rho +}^{K_{\rho +}/2}u_{\sigma +}^{K_{\sigma
+}/2}u_{\rho -}^{K_{\rho -}/2}u_{\sigma -}^{K_{\sigma -}/2}$, $X_{2}=u_{\rho
+}^{K_{\rho +}/2}u_{\sigma +}^{K_{\sigma +}/2}u_{\rho -}^{1/2K_{\rho
-}}u_{\sigma -}^{1/2K_{\sigma -}}$ and $j=\rho $ or $\sigma $. The initial
conditions for the above RG equations are as follows, $\mathcal{D}%
_{i}(0)=D_{i}\tilde{a}/(\pi v^{2})$, $K_{\rho +}(0)=u_{\rho +}^{-1}=K_{\rho
+}$, and $K_{\sigma +}(0)=K_{\rho -}(0)=K_{\sigma -}(0)=u_{\sigma
+}(0)=u_{\rho -}(0)=u_{\sigma -}(0)=1$. 
\begin{figure}[tb]
\begin{center}
\includegraphics[width=.5\textwidth,keepaspectratio=true]{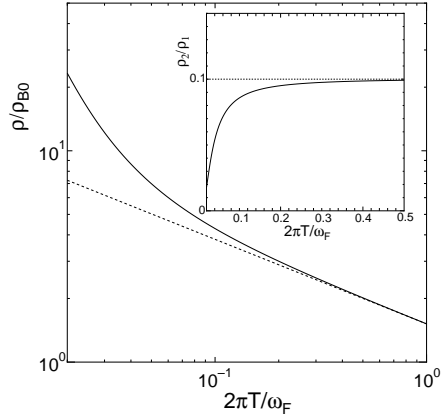}
\end{center}
\caption{ Temperature dependence of the resistivity $\protect\rho $ of SWNT
with the following parameters, $\protect\pi \protect\omega _{F}\protect\tau
_{1}=300$ and $\protect\pi \protect\omega _{F}\protect\tau _{2}=600$. The
solid (dotted) line corresponds to RG analysis (perturbation theory). Inset
: The ratio $\protect\rho _{2}/\protect\rho _{1}$ as a function of
temperature.}
\label{figresis}
\end{figure}
\qquad \qquad \qquad

The temperature dependence of the resistivity, Fig.~\ref{figresis}, can be
evaluated by solving the RG equations. The enhancement of the resistivity at
low-temperatures is due to the electron localization. Solution of RG
equations shows that the intra-valley (inter-valley) scattering pins the
phases, $\theta _{\rho +}$, $\theta _{\sigma +}$, $\theta _{\rho -}$, and $%
\theta _{\sigma -}$ ($\theta _{\rho +}$, $\theta _{\sigma +}$, $\phi _{\rho
-}$, and $\phi _{\sigma -}$). Since the conjugate variables, $\theta _{\rho
-}$ and $\phi _{\rho -}$, or $\theta _{\sigma -}$ and $\phi _{\sigma -}$
cannot be pinned at the same time, the localization due to two kinds of the
scattering cannot occur simultaneously. For the parameters of Fig. \ref
{figresis}, the intra-valley scattering dominates over the inter-valley
scattering at low temperatures (see inset of Fig. \ref{figresis}).

The high frequency behavior of the optical conductivity is not modified by
the effects of the localization. The power law behavior at low frequencies
survives due to a finite density of states at the Fermi energy (similarly to
the non-interacting case \cite{Berezinskii}).

\section{Effect of interactions beyond the Luttinger model}

\label{beyondluttinger}

Until now the Coulomb interaction in SWNTs has been treated within the
Luttinger model. In order to describe the interaction effects beyond this
approximation, one has to consider the full Hamiltonian (\ref{Hbos}). The
low energy properties of Eq. (\ref{Hbos}) can be investigated by the RG
method (see Ref. \cite{Giamarchi-Schulz} for details). At half-filling, $%
q_{F}=0$, we obtain the following RG equations, 
\begin{eqnarray}
(K_{\rho +})^{\prime } &=&-({K_{\rho +}^{2}}/{8}%
)(y_{1}^{2}+y_{2}^{2}+y_{3}^{2}+y_{7}^{2})\;\;,  \label{RG1} \\
(K_{\sigma +})^{\prime } &=&-({K_{\sigma +}^{2}}/{8}%
)(y_{1}^{2}+y_{5}^{2}+y_{6}^{2}+y_{8}^{2})\;\;,  \label{RG2} \\
(K_{\rho -})^{\prime } &=&-({K_{\rho -}^{2}}/{8}%
)(y_{2}^{2}+y_{4}^{2}+y_{6}^{2}+y_{9}^{2})\;\;,  \label{RG3} \\
(K_{\sigma -})^{\prime } &=&-({K_{\sigma -}^{2}}/{8}%
)(y_{3}^{2}+y_{4}^{2}+y_{5}^{2})+({1}/{8})(y_{7}^{2}+y_{8}^{2}+y_{9}^{2})\;%
\;,  \label{RG4} \\
(y_{1})^{\prime } &=&\left\{ 2-(K_{\rho +}+K_{\sigma +})\right\}
y_{1}-(y_{2}y_{6}+y_{3}y_{5}+y_{7}y_{8})/4\;\;,  \label{RG5} \\
(y_{2})^{\prime } &=&\left\{ 2-(K_{\rho +}+K_{\rho -})\right\}
y_{2}-(y_{1}y_{6}+y_{3}y_{4}+y_{7}y_{9})/4\;\;,  \label{RG6} \\
(y_{3})^{\prime } &=&\left\{ 2-(K_{\rho +}+K_{\sigma -})\right\}
y_{3}-(y_{1}y_{5}+y_{2}y_{4})/4\;\;,  \label{RG7} \\
(y_{4})^{\prime } &=&\left\{ 2-(K_{\rho -}+K_{\sigma -})\right\}
y_{4}-(y_{2}y_{3}+y_{5}y_{6})/4\;\;,  \label{RG8} \\
(y_{5})^{\prime } &=&\left\{ 2-(K_{\sigma +}+K_{\sigma -})\right\}
y_{5}-(y_{1}y_{3}+y_{4}y_{6})/4\;\;,  \label{RG9} \\
(y_{6})^{\prime } &=&\left\{ 2-(K_{\sigma +}+K_{\rho -})\right\}
y_{6}-(y_{1}y_{2}+y_{4}y_{5}+y_{8}y_{9})/4\;\;,  \label{RG10} \\
(y_{7})^{\prime } &=&\left\{ 2-(K_{\rho +}+1/K_{\sigma -})\right\}
y_{7}-(y_{1}y_{8}+y_{2}y_{9})/4\;\;,  \label{RG11} \\
(y_{8})^{\prime } &=&\left\{ 2-(K_{\sigma +}+1/K_{\sigma -})\right\}
y_{8}-(y_{1}y_{7}+y_{6}y_{9})/4\;\;,  \label{RG12} \\
(y_{9})^{\prime } &=&\left\{ 2-(K_{\rho -}+1/K_{\sigma -})\right\}
y_{9}-(y_{2}y_{7}+y_{6}y_{8})/4\;\;.  \label{RG13}
\end{eqnarray}
The initial conditions for Eqs. (\ref{RG1})-(\ref{RG13}) are $K_{j\delta
}(0)=K_{j\delta }$, $y_{1}=[\Delta V(0)-V_{pp}(2K_{0})]/(\pi v)$, $%
y_{2}=-y_{3}=-y_{5}=y_{6}=-\Delta V(0)/(\pi v)$, $y_{4}=[V_{pp}(2K)-\Delta
V(0)]/(\pi v)$, $y_{7}=-y_{8}=-V_{pp}(2K)/(\pi v)$, and $y_{9}=V_{pp}(2K)/(%
\pi v)$. In deriving the RG equations, the non-linear term $\cos 2\theta
_{\sigma -}\cos 2\phi _{\sigma -}$ is omitted because this operator stays
exactly marginal in all orders and is thus decoupled from the problem \cite
{Egger-Gogolin}. The RG equations away from half-filling can be obtained
from Eqs. (\ref{RG1})-(\ref{RG13}) by putting $y_{1}$, $y_{2}$, $y_{3}$ and $%
y_{7}$ to zero. Hereafter we concentrate on the case $N=10$, $\kappa =1.4$, $%
R_{s}$ = 100 $nm$ and $a_{0}/a=0.526$ and estimate the initial values of the
RG parameters using Table 1.

Away from half-filling, the quantities $K_{\sigma +}$, $K_{\rho -}$, and $%
K_{\sigma -}^{-1}$ renormalize to zero and the coefficient of $\cos 2\theta
_{\sigma +}\cos 2\theta _{\rho -}$ ($\cos 2\theta _{\sigma +}\cos 2\phi
_{\sigma -}$ and $\cos 2\theta _{\rho -}\cos 2\phi _{\sigma -}$) tends to $%
-\infty $ ($\infty $). As a result, the phases $\theta _{\sigma +},\theta
_{\rho -}$ and $\phi _{\sigma -}$ are pinned at $(\theta _{\sigma +},\theta
_{\rho -},\phi _{\sigma -})=(0,0,\pi /2)$ or $(\pi /2,\pi /2,0)$ so that the
modes $\sigma \pm $ and $\rho -$ are gapped. In this case, the asymptotic
behavior of the correlation functions at $x\rightarrow \infty $ is
determined by the correlations of the gapless $\rho +$ mode, $\left\langle 
\mathrm{e}^{\mathrm{i}n\theta _{\rho +}(x)}\mathrm{e}^{-\mathrm{i}n\theta
_{\rho +}(0)}\right\rangle \sim x^{-n^{2}K_{\rho +}/2}$ and $\left\langle 
\mathrm{e}^{\mathrm{i}m\phi _{\rho +}(x)}\mathrm{e}^{-\mathrm{i}m\phi _{\rho
+}(0)}\right\rangle \sim x^{-m^{2}/2K_{\rho +}}$ ($n=1$ and $2$ correspond
to $2q_{F}$ and $4q_{F}$ density waves and $m=1$ to a superconducting
state). Since $K_{\rho +}\approx 0.2$, the $2q_{F}$ density wave
correlations seem to be dominant. However, we found that the correlation
functions of any $2q_{F}$ density wave decay exponentially at large
distances due to the gapped modes. We therefore are looking for the
four-particle correlations. The $4q_{F}$ density waves dominate over
superconductivity for $K_{\rho +}<1/2$ \cite{Nagaosa,Schulz}. Such density
wave states are given by the product of the charge $\rho _{+}(x)\rho _{-}(x)$
or spin $S_{+}(x)S_{-}(x)$ densities at different sublattices. Substituting
the values of the pinned phases we observe that $\rho _{+}(x)\rho _{-}(x)$
vanishes, and the dominant state is the $4q_{F}$ spin density wave with
correlation function $\left\langle
S_{+}(x)S_{-}(x)S_{+}(0)S_{-}(0)\right\rangle \sim \cos 4q_{F}x/x^{2K_{\rho
+}}$.

At half-filling the solution of the RG equations (\ref{RG1})-(\ref{RG13})
indicates that the phase variables $\theta _{\rho +}$, $\theta _{\sigma +}$, 
$\theta _{\rho -}$, and $\phi _{\sigma -}$ are pinned and the all kinds of
excitation are gapped. In other words, the ground state of the half-filled
AN is a Mott insulator with spin gap. The same conclusion has been drawn
from the model with short range interactions \cite
{Balents-Fisher,Krotov-Lee-Louie}. The pinned phases are given by $(\theta
_{\rho +},\theta _{\sigma +},\theta _{\rho -},\phi _{\sigma -})=(0,0,0,0)$
or $(\pi /2,\pi /2,\pi /2,\pi /2)$ since the first, second and 6-9-th
non-linear terms in Eq. (\ref{Hbos}) scale to $-\infty $. The averages $%
\langle \rho _{+}(x)\rho _{-}(x)\rangle $ and $\langle
S_{+}(x)S_{-}(x)\rangle $ are both finite, which indicates formation of
bound states of electrons at different atomic sublattices.

The states derived from the present analysis are characteristic for the long
range Coulomb interaction. In fact, for the on-site plus nearest neighbor
interaction the dominant states correspond to the density waves at
half-filling and to the superconducting state or the density waves away from
half-filling \cite{Krotov-Lee-Louie}.

\section{Mott-insulating phase}

To estimate the gaps in different modes of excitations at half-filling, we
employ the self-consistent harmonic approximation which follows from
Feynman's variational principle \cite{Feynman}. We consider a trial harmonic
Hamiltonian of the form: 
\begin{eqnarray}
H_{0}=\sum_{j\delta }\frac{v_{j\delta }}{2\pi }\int \mathrm{d}x^{\prime }
&\{&K_{j\delta }^{-1}[(\partial _{x^{\prime }}\theta _{j\delta
})^{2}+(1-\delta _{j\sigma }\delta _{\delta -})q_{j\delta }^{2}\theta
_{j\delta }^{2}]  \nonumber \\
&+&K_{j\delta }[(\partial _{x^{\prime }}\phi _{j\delta })^{2}+\delta
_{j\sigma }\delta _{\delta -}q_{j\delta }^{2}\phi _{j\delta }^{2}]\},
\label{Hscha}
\end{eqnarray}
$q_{j\delta }$ being variational parameters. By minimizing the upper
estimate for the free energy $F^{\ast }=F_{0}+\left\langle
H-H_{0}\right\rangle _{0}$ one obtains the following self-consistent
equations, 
\begin{eqnarray}
q_{\rho +}^{2} &=&\frac{2K_{\rho +}}{\pi \tilde{a}^{2}v_{\rho +}}c_{\rho
+}[V_{pp}(2K)-\Delta V(0)]c_{\sigma +}+\Delta V(0)c_{\rho
-}+V_{pp}(2K)d_{\sigma -},  \label{qrp} \\
q_{\rho -}^{2} &=&\frac{2K_{\rho -}}{\pi \tilde{a}^{2}v_{\rho -}}c_{\rho
-}\Delta V(0)c_{\rho +}+\Delta V(0)c_{\sigma +}-V_{pp}(2K)d_{\sigma -},
\label{qrm} \\
q_{\sigma +}^{2} &=&\frac{2K_{\sigma +}}{\pi \tilde{a}^{2}v_{\sigma +}}%
c_{\sigma +}[V_{pp}(2K)-\Delta V(0)]c_{\rho +}+\Delta V(0)c_{\rho
-}-V_{pp}(2K)d_{\sigma -},  \label{gsp} \\
q_{\sigma -}^{2} &=&\frac{2}{\pi \tilde{a}^{2}K_{\sigma -}v_{\sigma -}}%
d_{\sigma -}V_{pp}(2K)\left\{ c_{\rho +}-c_{\sigma +}-c_{\rho -}\right\} ,
\label{qsm}
\end{eqnarray}
where $c_{j\delta }=\left\langle \cos 2\theta _{j\delta }\right\rangle
_{0}=\cos 2\theta _{j\delta }^{(m)}(\gamma \tilde{a}q_{j\delta
})^{K_{j\delta }}$, $d_{\sigma -}=\left\langle \cos 2\phi _{\sigma
-}\right\rangle _{0}=\cos 2\phi _{\sigma -}^{(m)}(\gamma \tilde{a}q_{\sigma
-})^{1/K_{\sigma -}}$, $\left\langle ...\right\rangle _{0}$ denotes
averaging with respect to the trial Hamiltonian (\ref{Hscha}), and $\gamma
\simeq 0.890$. Note that $\left\langle \cos 2\theta _{\sigma -}\right\rangle
_{0}=0$, so that only the terms of the Hamiltonian (\ref{Hbos}) which scale
to the strong coupling (see end of Section \ref{beyondluttinger}) contribute
to Eqs. (\ref{qrp})-(\ref{qsm}).

In the limiting case of interest, $|\Delta V(0)|,|V_{pp}(2K)|\ll v$ and $%
K_{\rho +}\ll 1$, the solution of Eqs. (\ref{qrp})-(\ref{qsm}) can be found
in a closed form, giving rise to the following estimates for the gaps $%
\Delta _{j\delta }=v_{j\delta }q_{j\delta }$ in the energy spectra, 
\begin{eqnarray}
\Delta _{\rho +} &=&\frac{v_{\rho +}}{\gamma \tilde{a}}\left( \frac{2\gamma
^{2}V_{\rho +}}{\pi v_{\rho +}}\right) ^{1/(1-K_{\rho +})}  \label{drp} \\
\Delta _{\rho -} &=&\frac{|\Delta V(0)|}{V_{\rho +}}\Delta _{\rho +}
\label{drm} \\
\Delta _{\sigma +} &=&\frac{|V_{pp}(2K_{0})-\Delta V(0)|}{V_{\rho +}}\Delta
_{\rho +}  \label{dsp} \\
\Delta _{\sigma -} &=&\frac{|V_{pp}(2K_{0})|}{V_{\rho +}}\Delta _{\rho +}
\label{dsm}
\end{eqnarray}
with $V_{\rho +}=\left\{ [\Delta V(0)-V_{pp}(2K)]^{2}+[\Delta
V(0)]^{2}\right. $ $\left. +[V_{pp}(2K)]^{2}\right\} ^{1/2}$. In the above
expressions we used the approximation, $v/v_{\rho +}=K_{\rho +}$ and $%
v/v_{j\delta }=K_{j\delta }=1$ for $\sigma \pm $ and $\rho -$ modes. The
formulae (\ref{drp})-(\ref{dsm}) indicate that the largest gap occurs in the 
$\rho +$ mode, albeit all four gaps are of the same order for realistic
values of the matrix elements (see Table 1). The gaps decrease as $\Delta
_{j\delta }\propto (1/R)^{1/(1-K_{\rho +})}\simeq 1/R^{5/4}$ with the
nanotube radius. 
\begin{figure}[tb]
\begin{center}
\includegraphics[width=.5\textwidth]{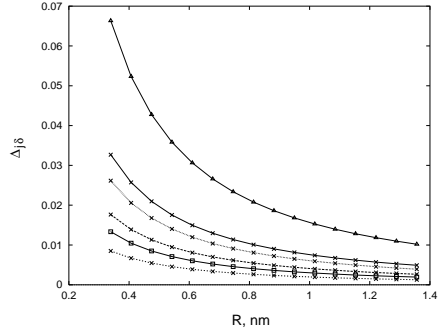}
\end{center}
\caption{The energy gaps $\Delta _{j\protect\delta }$ for the modes $\protect%
\rho +$, $\protect\sigma -$, $\protect\sigma +$, $\protect\rho -$ at $%
a_{0}=0.526a$ (lines marked by crosses, from top to bottom) and for the mode 
$\protect\rho +$ at $a_{0}=0.4a$ (triangles) and at $a_{0}=0.7a$ (squares).
The energy is in units $\hbar v/\tilde{a}\simeq 2.16$ eV for $\tilde{a}=a$.}
\label{figgap}
\end{figure}

In Figure \ref{figgap} we present numerical results for the gaps $\Delta
_{j\delta }$ for the short distance cutoff of the Coulomb interaction $%
a_{0}=0.526a$. The data for somewhat larger and somewhat smaller values of $%
a_{0}$ indicate possible variations of the gap $\Delta _{\rho +}$ due to the
uncertainty in the cutoff. The gaps can be roughly estimated at $\Delta
_{j\delta }\sim 0.01-0.1$ eV for typical SWNTs with $R\simeq 0.7$ nm.

The resistivity $\rho $ of metallic SWNTs increases exponentially $\rho
\propto \exp (\Delta _{\rho +}/T)$ at low temperatures $T\ll \Delta _{\rho +}
$. On the other hand, at high temperatures, $T\gg \Delta _{\rho +}$,
perturbation theory with respect to the non-linear terms of the Hamiltonian (%
\ref{Hbos}) gives a power-law behavior of the resistivity $\rho \sim
T^{2K_{\rho +}-1}/N^{2}$ due to umklapp scattering at half-filling. Note
that the power-law is different from that governing the impurity scattering,
Eq. (\ref{rhoimpurity}). A resonant increase of the resistivity of SWNTs at
half-filling is a characteristic signature of the Mott insulating phase.

Due to the gaps in the spectrum of bosonic elementary excitations, the
electronic density of states should disappear in the subgap region and
display features at the gap frequencies and their harmonics. These
signatures should be observable by means of the tunneling spectroscopy.

\section{Observability of the Mott-insulating phase}

In this section we discuss various factors which might influence the
observability of the Mott-insulating phase in realistic systems. An
exponential increase of the resistivity of half-filled metallic SWNTs at low
temperatures can also occur due to deformation-induced gaps $\Delta _{d}$\
in the single-particle spectrum \cite{Kane-Mele}. These gaps are in the
range $\Delta _{d}<0.02$ eV for typical SWNTs with $R\simeq 0.7$ nm. This
value is comparable with (but somewhat smaller than) our estimate for the
Mott gaps. We therefore cannot exclude the interplay between the two
mechanisms suppressing electron transport at half-filling. Note that the
deformation induced gaps decrease as $1/R^{2}$ with increasing nanotube
radius \cite{Kane-Mele}, which should be contrasted to $1/R^{5/4}$
dependence for the Mott gaps. Moreover, deformation-induced gaps strongly
depend on the chiral angle of SWNTs, whereas Mott gaps are determined by the
radius of SWNTs and depend weakly on their chirality.

Isolated metallic SWNTs are half-filled systems. However, in generic
experimental layouts the nanotubes are contacted to metallic electrodes. The
difference $\Delta W$ in the work functions of a metal (typically Au or Pt)
and nanotube results in a charge transfer between the nanotube and
electrodes, which shifts the Fermi level of the nanotube downwards away from
half-filling (for $\Delta W>0$).

\begin{figure}[tb]
\begin{center}
\includegraphics[width=.5\textwidth]{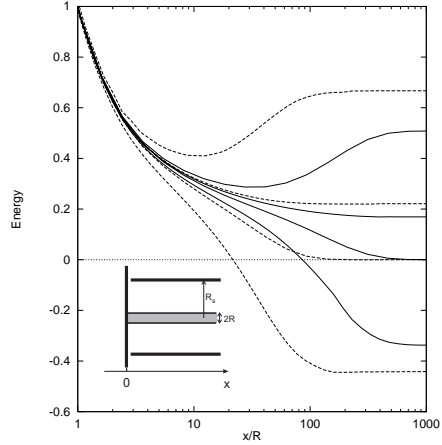}
\end{center}
\caption{The energy $\Delta E$ of the gapless point (half filling) with
respect to the Fermi level (units of $\Delta W/(2e^2 \protect\nu/\protect%
\kappa)$). The curves from top to bottom at $x/R=1000$ correspond to $%
eV_{g}/\Delta W=-2,0,1,3$. The screening radius $R_{s}/R=75$, $300$ for
dashed and solid curves corresponds approximately to $50$ nm and $200$ nm
for (10,10) SWNTs. The Coulomb interaction is characterized by $\protect\nu 
\bar{V}(0) / \ln (R_{s}/R) = 5$. Inset: layout of the system. }
\label{figchneutr}
\end{figure}

To be specific, consider a metallic SWNT surrounded by a coaxial cylindrical
gate electrode of radius $R_{s}\gg R$. The nanotube ($x>0$) contacts the $yz$%
-plane of metallic electrode ($x<0$) at $x=0$ (see inset of Fig. \ref
{figchneutr}). The interaction term of the Hamiltonian (\ref{Hrhoplus}) is
modified by taking the electrostatic potential $V_{g}$ of the gate as well
as the image charge at the metallic electrode into account, 
\begin{eqnarray}
H_{int} &=&\int_{0}^{\infty }dxdx^{\prime }\{\frac{1}{2}\rho
(x)[V(x-x^{\prime })-V(x+x^{\prime })]\rho (x^{\prime })  \label{HintMNT} \\
&&+\rho (x)M(x-x^{\prime })eV_{g}sign(x^{\prime })\}+\Delta
W\int_{0}^{\infty }dx\rho (x),  \nonumber
\end{eqnarray}
The Fourier images of the kernels $V(x)$, $M(x)$ are given by, 
\begin{equation}
\bar{V}_{q}=\frac{2e^{2}}{\kappa }\left\{ I_{0}(qR)K_{0}(qR)-I_{0}^{2}(qR)%
\frac{K_{0}(qR_{s})}{I_{0}(qR_{s})}\right\} ,\ \bar{M}_{q}=\frac{I_{0}(qR)}{%
I_{0}(qR_{s})},  \label{Uq}
\end{equation}
with the modified Bessel functions $I_{0}$, $K_{0}$. The kernel $\bar{V}_{q}$
describes the long-range Coulomb interaction, $V(x)=1/\kappa |x|$, for $R\ll
|x|\ll R_{s}$. The interaction is screened at large distances $|x|\gg R_{s}$%
, so that $\bar{V}_{q=0}=(2e^{2}/\kappa )\ln (R_{s}/R)$, in agreement with
our previous estimate (see text below Eq. (\ref{Hint})).

Minimization of the total energy given by the Hamiltonian (\ref{Hrhoplus})
with the interaction term (\ref{HintMNT}) determines the profile of the
charge density $e\rho (x)$ and the deviation $-\Delta E(x)$ of the Fermi
level from half-filling, $\Delta E=\rho /\nu $ (see Refs. \cite{OTpb,OTjltp}%
\ for details), 
\begin{eqnarray}
\Delta E(x) &=&\frac{\Delta W}{\nu \bar{V}(0)}\frac{\ln (R_{s}/R)}{\ln (z/R)}%
-\frac{ceV_{g}}{\nu \bar{V}(0)}\frac{x}{R_{s}},\ \mbox{for }R\ll x\ll R_{s},
\label{E0smallz} \\
\Delta E(x) &=&\frac{\Delta W-eV_{g}}{1+\nu \bar{V}(0)},\ \mbox{for }%
R_{s}\ll x,  \label{E0largez}
\end{eqnarray}
where $c=(1/\pi )\int dx|I_{0}(x)|^{-1}\simeq 1.33$ and the Coulomb
interaction is supposed to be strong, $\nu \bar{V}(0)\gg \ln (R_{s}/R)$.

Equation (\ref{E0smallz}) shows that the density of charge transferred to
SWNT due to the mismatch $\Delta W$ of the work-functions decays
logarithmically slow with the distance $x$\ from the metallic electrode \cite
{OTpb}. This is related to the poor screening of the Coulomb interaction in
1D SWNTs. The influence of a gate voltage on the charge density in SWNT is
suppressed by a factor $x/R_{s}$ in the vicinity of electrode, $x<R_{s}$,
cf. Eqs. (\ref{E0smallz}), (\ref{E0largez}). In order to achieve
half-filling condition $\Delta E=0$ in samples with small distances $d<R_{s}$
between the electrodes, one should compensate for this effect by applying a
higher gate voltage.

The coordinate dependence of the charge density (or the deviation $\Delta
E(x)$ from half-filling) is shown in Fig. \ref{figchneutr} for several
values of the gate voltage and the screening radius. If a substantial
positive voltage is applied to the gate, the charge density in SWNT can
change sign (from positive to negative) with increasing distance from the
metallic electrode. Let us note that the Mott gap should result in the
formation of half-filled incompressible regions near such ''charge
neutrality points'', $\Delta E(x)=0$. This is in contrast to isolated SWNTs
where the Mott transition occurs uniformly in the whole system. The electron
transport through the incompressible regions seems to be an interesting
problem for future research.

\section{Conclusions}

We have investigated manifestations of electron correlations in metallic
single-wall carbon nanotubes. Effective low-energy model of metallic SWNTs
with arbitrary chirality is developed starting from microscopic
considerations. The unrenormalized parameters of the model show very weak
dependence on the chiral angle, which makes the low-energy properties of
metallic SWNTs chirality-independent. In the absence of inter-branch
electron scattering the model corresponds to two-channel Luttinger liquid.
The impurity scattering in such Luttinger liquid is investigated using
memory function technique and the renormalization group method. We show that
the intra-valley and inter-valley backscattering can not coexist at low
energies. The properties of clean SWNTs are investigated beyond the
Luttinger liquid limit. The ground state away from half-filling is found to
be $4q_{F}$ spin density wave. At half-filling the umklapp scattering
affects strongly interacting mode of total charge excitations. This makes
the umklapp scattering strongly relevant perturbation which gives rise to
the Mott-insulating ground state. The latter is characterized by ''binding''
of electrons at two atomic sublattices of graphite. The energy gaps in all
modes of excitations are evaluated within self-consistent harmonic
approximation. Finally, observability of the Mott phase in realistic SWNTs
is discussed.

\section{Acknowledgments}

The authors would like to thank B.L. Altshuler, G.E.W. Bauer, C. Dekker,
Yu.V. Nazarov, S.Tans, S. Tarucha, Y. Tokura, and N. Wingreen for
stimulating discussions. Financial support of the Royal Dutch Academy of
Sciences (KNAW) is gratefully acknowledged.

%


\begin{thebibliography}{99}
\bibitem{DekkerRev}  For a recent review see C. Dekker, Physics Today 
\textbf{5}, 22 (1999)

\bibitem{Tans}  S.J. Tans, M. H. Devoret, H. Dai, A. Thess, R. E. Smalley,
L. J. Geerligs, C. Dekker\textit{, } Nature (London) \textbf{386}, 474,
(1997)

\bibitem{Bockrath}  M. Bockrath, D. H. Cobden, B. L. McEuen, N. G. Chopra,
A. Zettl, A. Thess, R. E. Smalley\textit{, } Science, \textbf{275}, 1922
(1997)

\bibitem{Wildoer}  J.W.G. Wild\"{o}er, L.C. Venema, A.G. Rinzler, R.E.
Smalley, C. Dekker\textit{, } Nature (London) \textbf{391}, 59 (1998)

\bibitem{Odom}  T.W. Odom, J. Huang, P. Kim, C.M. Lieber, Nature (London) 
\textbf{391}, 62 (1998)

\bibitem{Johnson}  W. Clauss, D. J. Bergeron, A. T. Johnson, Phys. Rev. B 
\textbf{58}, R4266 (1998)

\bibitem{Venema}  L.C. Venema, J.W.G. Wildoer, J.W. Janssen, S.J. Tans, H.
Tuinstra, L.P. Kouwenhoven, C. Dekker\textit{, }Science \textbf{283}, 52
(1999)

\bibitem{BockrathLL}  M. Bockrath, D.H. Cobden, J. Lu, A.G. Rinzler, R.E.
Smalley, L. Balents, and P.L. McEuen\textit{,} Nature (London) \textbf{397},
598 (1998)

\bibitem{SchonenbergerLL}  C. Schon\"{e}nberger, A. Bachtold, C. Strunk,
J.-P. Salvetat, L. Forro, Appl. Phys. A 69, \textbf{283} (1999)

\bibitem{Yao}  Z. Yao, H. Postma, L. Balents, and C. Dekker,
Nature (London) \textbf{402}, 273 (1999)


\bibitem{Fuhrer}  M.S. Fuhrer, J. Nyg\aa rd, L. Shih, M. Bockrath, A. Zettl,
and P. McEuen\textit{,} submitted to Nature (London)

\bibitem{Tans3}  S.J. Tans, A.R.M. Verschueren, and C. Dekker, Nature
(London) \textbf{393}, 49 (1998)

\bibitem{KBF}  C. Kane, L. Balents and M. P. A. Fisher, Phys. Rev. Lett. 
\textbf{79}, 5086 (1997)

\bibitem{Egger-Gogolin}  R. Egger and A. O. Gogolin, Phys. Rev. Lett. 
\textbf{79}, 5082 (1997); Eur. Phys. J. B \textbf{3}, 281 (1998)

\bibitem{YOprl99}  H. Yoshioka and A.A. Odintsov, Phys. Rev. Lett., \textbf{%
82}, 374 (1999)

\bibitem{Yoshioka-impurities}  H. Yoshioka, preprint cond-mat/9903342

\bibitem{OYprb99}  A.A. Odintsov and H. Yoshioka, Phys. Rev. B \textbf{59},
R10457 (1999)

\bibitem{Wallace}  P. R. Wallace, Phys. Rev. \textbf{71}, 622 (1947)

\bibitem{Moore}  see e.g. E. Moore, B. Gherman, and D. Yaron, J. Chem. Phys. 
\textbf{106}, 4216 (1997)

\bibitem{Lin-Balents-Fisher}  H. Lin, L. Balents, and M.P.A. Fisher, Phys.
Rev. B \textbf{58}, 1794 (1998)

\bibitem{Ando-Nakanishi}  T. Ando and T. Nakanishi, J. Phys. Soc. Jpn. 
\textbf{67}, 1704 (1998)

\bibitem{memory}  H. Mori, Prog. Theor. Phys. \textbf{33}, 423 (1965) 
\textbf{34}, 399 (1965); W. G\"{o}tze and P. W\"{o}lfle, Phys. Rev. B 
\textbf{6}, 1126 (1972)

\bibitem{Giamarchi-Schulz}  T. Giamarchi and H. J. Schulz, Phys. Rev. B 
\textbf{37}, 325 (1988)

\bibitem{Berezinskii}  V. L. Berezinski$\breve{\mathrm{i}}$, Sov. Phys.
JETP. \textbf{38}, 620 (1974)

\bibitem{Nagaosa}  N. Nagaosa, Solid State Communications \textbf{94}, 495
(1995)

\bibitem{Schulz}  H. J. Schulz, Phys. Rev. B \textbf{53}, R2959 (1996)

\bibitem{Balents-Fisher}  L. Balents and M. P. A. Fisher, Phys. Rev. B 
\textbf{55}, R11973 (1997)

\bibitem{Krotov-Lee-Louie}  Yu. A. Krotov, D.-H. Lee, and Steven G. Louie,
Phys. Rev. Lett. \textbf{78}, 4245 (1997)

\bibitem{Feynman}  R.P. Feynman, \textit{Statistical mechanics}, Reading,
Mass., W.A. Benjamin, Inc., 1972

\bibitem{Kane-Mele}  C.L. Kane and E.J. Mele, Phys. Rev. Lett. \textbf{78},
1932 (1997)

\bibitem{OTpb}  A.A. Odintsov and Y. Tokura, to be published in \emph{%
Proceedings of the LT-22, Helsinki, 1999}, preprint cond-mat/9906269

\bibitem{OTjltp}  A.A. Odintsov and Y. Tokura, to be published.
\end{thebibliography}
\end{document}